**Synthesis of a High-Capacity α-Fe$_2$O$_3$@C Conversion Anode and a High-Voltage LiNi$_{0.5}$Mn$_{1.5}$O$_4$ Spinel Cathode and their Combination in a Li-Ion Battery**

Shuangying Wei[a], Daniele Di Lecce[b], Riccardo Messini D'Agostini[a], Jusef Hassoun[a,b,c*]

[a] *Department of Chemical, Pharmaceutical and Agricultural Sciences, University of Ferrara, Via Fossato di Mortara, 17, 44121, Ferrara, Italy.*

[b] *Graphene Labs, Istituto Italiano di Tecnologia, via Morego 30, Genova, 16163, Italy*

[c] *National Interuniversity Consortium of Materials Science and Technology (INSTM), University of Ferrara Research Unit, Via Fossato di Mortara, 17, 44121, Ferrara, Italy.*

Corresponding author: jusef.hassoun@iit.it, jusef.hassoun@unife.it

**Abstract**

Li-conversion α-Fe$_2$O$_3$@C nanocomposite anode and high-voltage LiNi$_{0.5}$Mn$_{1.5}$O$_4$ cathode are synthesized in parallel, characterized, and combined in a Li-ion battery. The α-Fe$_2$O$_3$@C is prepared via annealing under argon of maghemite iron oxide and sucrose, and subsequent oxidation in air. The nanocomposite exhibits a satisfactory electrochemical response in a lithium half-cell delivering almost 900 mAh g$^{-1}$, as well as significantly longer cycle life and higher rate capability compared to the bare iron oxide precursor. The LiNi$_{0.5}$Mn$_{1.5}$O$_4$ cathode, achieved using a modified co-precipitation approach, reveals well defined spinel structure without impurities, a sub-micrometrical morphology, and a reversible capacity of ca. 120 mAh g$^{-1}$ in a lithium half-cell with an operating voltage of 4.8 V. Hence, a lithium-ion battery is assembled by coupling the α-Fe$_2$O$_3$@C anode with the LiNi$_{0.5}$Mn$_{1.5}$O$_4$ cathode. This cell operates at about 3.2 V, delivering a stable capacity of 110 mAh g$^{-1}$ (referred to the cathode mass) with coulombic efficiency exceeding 97%. Therefore, this cell is suggested as promising energy storage system with expected low economic and environmental impact.

**Introduction**

Electric vehicles (EVs), hybrid-electric vehicles (HEVs) and plug-in hybrid electric vehicles (PHEVs) are predominantly powered by the most common version of the lithium-ion battery, i.e., the one combining a graphite anode with a layered transition metal oxide cathode and employed in common portable electronics.[1–4] This system is based on the electrochemical (de)insertion of lithium into and from the electrode materials[5] and can typically store ca. 250 Wh kg$^{-1}$ for a high number of charge/discharge cycles.[4,6] Graphite uptakes Li$^+$ delivering a capacity of 370 mAh g$^{-1}$, which is limited by an amount of alkali metal ions stored within the carbon layers reaching a maximum of 0.16 Li-equivalents per mole of C, i.e., according to the LiC$_6$ chemical formula.[7,8] Transition metal oxides react in the cell according to an electrochemical conversion pathway mainly occurring below 1.5 V vs Li$^+$/Li and involving a multiple exchange of electrons which ensures a higher capacity than that of graphite.[9–14] However, this intriguing class of materials intrinsically suffers from poor electrical conductivity and a large volume change throughout the electrochemical process, which cause voltage hysteresis and rapid cell decay upon cycling.[15] A suitable strategy to mitigate the various issues hindering an efficient use of these alternative anodes is represented by engineering nanostructured oxides with an increased active surface.[16] Indeed, Li-conversion materials, such as CuO, NiO, and MnO, can react with lithium to form the corresponding metal (Cu, Ni, and Mn) and lithium oxide (Li$_2$O), with the remarkably high specific capacity of 650,[17] 883,[18] and 440 mAh g$^{-1}$,[19] respectively. Higher capacity values, i.e., 910 mAh g$^{-1}$, may be achieved using α-Fe$_2$O$_3$, which react via the conversion pathway leading to Fe and Li$_2$O.[20,21] For instance, nanomembranes,[22] nanofibers,[23] nanowires,[24] and nanobelts [25] of α-Fe$_2$O$_3$ have been synthesized to increase the cycle life and the rate capability of the cell. However, this approach was limited by possible undesired reactions between the electrolyte and the oxide nanoparticles, which lead to capacity decay upon cycling or even to safety concerns,[26] as well as by the relevant costs due to complex synthesis techniques.[27,28] Highly conductive carbon matrixes incorporating α-Fe$_2$O$_3$ have been proposed as a possible pathway to decrease the hysteresis of the conversion reaction[29] and achieve *ad hoc* designed nanostructures.[30] Indeed, γ-Fe$_2$O$_3$@CNTs,[31] Fe$_2$O$_3$@C flakes,[20] and Fe$_2$O$_3$@C nanocomposites[32] have shown improved electrochemical process and excellent cycling stability. In our previous work we have shown the cell performances of a Fe$_2$O$_3$–MCMB composite obtained by high-energy ball milling, which revealed satisfactory characteristics for application in lithium battery.[33] Besides, a fine material tuning[31] combined with simple synthetic approaches of iron oxide nanocomposites might actually facilitate practical applications and scaling up.[28,34] We have lately adopted a facile two-step synthesis to prepare a C-coated nanostructured NiO anode successfully used in a novel lithium-ion battery using the LiNi$_{1/3}$Co$_{1/3}$Mn$_{1/3}$O$_2$ layered cathode.[35] However, voltage hysteresis and a working voltage higher than that of conventional graphite may actually jeopardize the potential use of conversion electrodes in a practical full Li-ion cell.[35] On the other hand, high-voltage cathodes exploiting the spinel-type structure and a Li$_x$M$_y$N$_{(2-y)}$O$_4$ chemical formula (where M and N are transition metals, e.g., Ni, Mn, or Fe) appeared as the ideal candidates to enable the use of the Li-conversion anodes.[36–39] Among them, LiNi$_{0.5}$Mn$_{1.5}$O$_4$ revealed the most suitable performance in lithium cell, namely, a working voltage of 4.8 V, a theoretical capacity of 147 mAh g$^{-1}$, and high rate capability.[40–42] Herein we extended the approach previously adopted for the synthesis of NiO@C[35] to prepare a C-coated α-Fe$_2$O$_3$ nanocomposite anode and concomitantly prepared by *ad hoc* developed method a LiNi$_{0.5}$Mn$_{1.5}$O$_4$ spinel cathode. The α-Fe$_2$O$_3$@C has been obtained from maghemite iron oxide (γ-Fe$_2$O$_3$) and sucrose, by annealing in reducing conditions with subsequent oxidation at mild temperature in air. The LiNi$_{0.5}$Mn$_{1.5}$O$_4$ spinel cathode has been achieved using a modified co-precipitation pathway including acetate and oxalic acid, leading to sub-micrometric active particles. Structure, morphology, composition, and electrochemical behavior in lithium half-cells of these two electrodes have been investigated. Subsequently, the α-Fe$_2$O$_3$@C anode and LiNi$_{0.5}$Mn$_{1.5}$O$_4$ cathode have been coupled in a new lithium-ion full-cell efficiently operating at 3.2V.

**Experimental**

**Synthesis of α-Fe$_2$O$_3$@C**

3.0 g of γ-Fe$_2$O$_3$ (Sigma-Aldrich, 50 nm) and 3.0 g of sucrose (Sigma-Aldrich) were dispersed in a solution of water and ethanol in the 1:1 v/v ratio (50 mL), and stirred at 70 °C until solvent evaporation (ca. 6 h). The mixture was treated for 10 h at 120 °C in an argon atmosphere, and then annealed for 3 h at 700 °C (heating rate was 5 °C min$^{-1}$) to obtain a reduced sample (indicated as α-Fe@C), which was grinded and subsequently heated in air at 380 °C for 24 h (heating rate was 5 °C min$^{-1}$) to prepare the final composite (indicated as α-Fe$_2$O$_3$@C).

**Synthesis of LiNi$_{0.5}$Mn$_{1.5}$O$_4$**

LiNi$_{0.5}$Mn$_{1.5}$O$_4$ was prepared using a modified co-precipitation pathway.[43,44] LiCH$_3$COO·2H$_2$O (99%, Sigma-Aldrich), Ni(CH$_3$COO)$_2$·4H$_2$O (99%, Sigma-Aldrich), and Mn(CH$_3$COO)$_2$·4H$_2$O (99%, Sigma-Aldrich) were dissolved with a Li:Ni:Mn molar ratio of 1.1:0.5:1.5 in a water-ethanol mixture to get solution A (water:ethanol 1:5 v/v).

Furthermore, $H_2C_2O_4 \cdot 2H_2O$ ($C_2H_2O_4 \cdot 2H_2O$, 99%, Aldrich) was dissolved in an identical hydro-alcoholic solution (B). This latter solution (B) was dropwise added to solution A while stirring the mixture and then kept at ambient temperature for 12 h to precipitate the metal oxalates. Afterward, the precipitate was treated for 12 h at 80 °C under constant stirring to evaporate water and ethanol. The precipitate was annealed for 6 h at 500 °C in a dry air flow to obtain an oxide powder (heating ramp was 5 °C min$^{-1}$). This powder was ground in a mortar, pressed into pellets, and calcined for 12 h at 800 °C in a dry air flow with to obtain $LiNi_{0.5}Mn_{1.5}O_4$ (heating ramp was 5 °C min$^{-1}$).

**Characterization**

X-ray diffraction (XRD) patterns of the materials were collected using a Bruker D8-Advance equipped with a Cu Kα source by changing the 2θ angle with steps of 0.02° every 10 s. Scanning electron microscopy (SEM, Zeiss EVO 40) and transmission electron microscopy (TEM, Zeiss EM 910) analyses were performed to detect the morphology and microstructure of the powders. Sample composition was analyzed by energy dispersive X-ray spectroscopy (EDS), employing the X-ACT Cambridge Instrument analyzer of the above-mentioned scanning electron microscope.

Electrode slurries were prepared dispersing the active material, poly(vinylidene fluoride) (Solef® 6020 PVDF), and conductive Super P carbon black (Timcal) in N-methyl pyrrolidone (NMP, Sigma-Aldrich), with a weight ratio between the solid components of by 8:1:1. This slurry was cast on copper (for α-$Fe_2O_3$@C, α-Fe@C, and γ-$Fe_2O_3$) or aluminum (for $LiNi_{0.5}Mn_{1.5}O_4$) foils by using a doctor blade (MTI Corporation). After NMP evaporation at ca. 70 °C, disks with a diameter of 10 and 14 mm were cut out from the electrode foils and kept overnight under vacuum at 110 °C.

The mass loading of α-Fe$_2$O$_3$@C and LiNi$_{0.5}$Mn$_{1.5}$O$_4$ over the electrodes was about 2.0 mg cm$^{-2}$ and 6.1 mg cm$^{-2}$, respectively.

The electrochemical performance was determined in half cells that used lithium metal disk as counter/reference electrode and a glass fiber separator (Whatman GF/A) soaking an 1 M electrolyte solution of LiPF$_6$ in EC:DMC 1:1 (v/v). Coin-type cells (CR2032, MTI Corporation) and T-type cells were made in a glovebox (Ar atmosphere, MBraun) where H$_2$O and O$_2$ contents were kept below 1 ppm. The behavior of the α-Fe$_2$O$_3$@C electrode in the lithium cell (as well as those of α-Fe@C and γ-Fe$_2$O$_3$ for comparison) was investigated by cyclic voltammetry (CV) and electrochemical impedance spectroscopy (EIS, T-type cell configuration) employing a VersaSTAT MC Princeton Applied Research (PAR) multichannel potentiostat. CV data were collected within the voltage range from 0.01 to 2.8 V *vs.* Li$^+$/Li using a scan rate of 0.1 mV s$^{-1}$. Impedance spectra were taken before cycling the cell (i.e., open circuit voltage, OCV) as well as after the 1st, 2nd, and 3rd cycles, using an alternate voltage with an amplitude of 10 mV from a frequency of 500 kHz to a frequency of 100 mHz. Galvanostatic cycling tests were carried out to study the electrodes in coin-type, lithium half-cells, which were charged and discharged between 0.01 and 2.8 V at C/5 (for α-Fe$_2$O$_3$@C, α-Fe@C and γ-Fe$_2$O$_3$) and between 3 and 5 V at C/2 (for LiNi$_{0.5}$Mn$_{1.5}$O$_4$). 1C was 1007 mA g$^{-1}$ for α-Fe$_2$O$_3$@C α-Fe@C and γ-Fe$_2$O$_3$, and 147 mA g$^{-1}$ for LiNi$_{0.5}$Mn$_{1.5}$O$_4$. A rate capability test of α-Fe$_2$O$_3$@C was carried out by changing stepwise the current from C/10 to 2C in the 0.01 − 2.8 V range.

A Li-ion cell was assembled coupling the α-Fe$_2$O$_3$@C anode with the LiNi$_{0.5}$Mn$_{1.5}$O$_4$ cathode. Before use in the full cell, the α-Fe$_2$O$_3$@C electrode was electrochemically treated in a half-cell by performing 3 galvanostatic cycles at C/5 between 0.01 and 2.8 V, with the last charge ending at 2.1 V, in order to ensure a steady working condition of the anode in the lithium-ion cell. The α-Fe$_2$O$_3$@C/LiNi$_{0.5}$Mn$_{1.5}$O$_4$ battery was made using a ratio between negative and positive electrode (i.e., N/P ratio) of about 2.2, as determined by taking into account the theoretical capacity of α-Fe$_2$O$_3$@C (1007 mAh g$^{-1}$) and LiNi$_{0.5}$Mn$_{1.5}$O$_4$ (147 mAh g$^{-1}$) along with their mass loading (2.0 and 6.1 mg cm$^{-2}$, respectively). This full-cell was charged and discharged at a C/2 rate in the 1.5 – 4.5 V voltage range (where 1C is 147 mA g$^{-1}$ as referred to the cathode mass). All galvanostatic measurements were conducted at room temperature with a Maccor Series 4000 battery tester.

**Results and discussion**

Figure 1 reports in a diagram the synthetic steps employed to prepare the α-Fe$_2$O$_3$@C nanocomposite (see details in the Experimental Section). The synthesis involves reduction of pristine γ-Fe$_2$O$_3$ with sucrose at 700 °C under argon to form an α-Fe core and a carbon shell, and subsequent oxidation to α-Fe$_2$O$_3$@C at 380 °C under air. This pathway, previously adopted for achieving a NiO@C electrode, leads to carbon-coated metal oxide particles suitable for battery application and is characterized by a simple experimental setup.[35] In addition, pristine γ-Fe$_2$O$_3$ and sucrose are low price and large available precursors, thus possibly providing a scalable two-step method with a modest economic impact to prepare of an efficient and alternative iron oxide anode for use in Li-ion battery.[28]

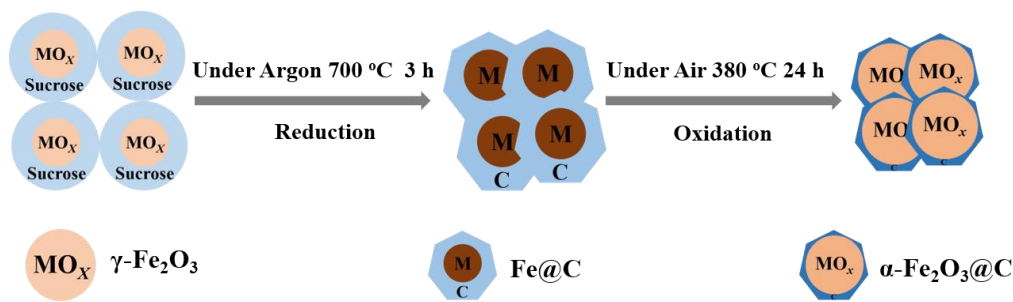

**Figure 1.** Illustration of the synthetic steps of α-Fe$_2$O$_3$@C (see details in Experimental Section).

Structure, morphology, and elemental composition of the α-Fe$_2$O$_3$@C material are provided in Figure 2. The XRD patterns of Fig. 2a indicate a cubic structure both for the precursor (γ-Fe$_2$O$_3$, JCPDS # 39-1346) and the synthesis intermediate (α-Fe, JCPDS # 06-0696), while the final material shows a hexagonal arrangement (α-Fe$_2$O$_3$, JCPDS # 33-0664). The above reflections reveal a substantial change of phase and crystal structure during the synthesis steps: indeed, the data suggest an almost complete reduction of the pristine maghemite (γ-Fe$_2$O$_3$) to metallic iron (α-Fe) by annealing under argon in presence of sucrose, and a subsequent oxidation to hematite (α-Fe$_2$O$_3$) upon the mild treatment under air. Therefore, the pure cubic phase of γ-Fe$_2$O$_3$ converts into α-Fe due to the reducing environment provided by high-temperature pyrolysis of sucrose into carbon, whilst the formation of α-Fe$_2$O$_3$ is promoted by direct reaction of iron metal with oxygen at 380 °C.[45] Relevantly, the patterns of the intermediate (α-Fe and pyrolytic C) and the final powder (α-Fe$_2$O$_3$ and C residue) do not show any reflections of graphite, expected at 2θ values of about 20, 40 and 60°,[7] thus suggesting the formation of lowly crystalline carbon in the α-Fe$_2$O$_3$@C. A further insight on the material characteristics is given by the SEM and TEM images of the bare precursor (Fig. 2b, d), the synthesis intermediate (Fig. 2e, g), and the final sample (Fig. 2h, j), respectively. The SEM images reveal that all materials are formed by the aggregation of submicron primary particles into secondary ones with a size ranging from 1 to 10 μm (Fig. 2d, e, h). Despite the similar morphology, the SEM image of the α-$Fe_2O_3$-C appears more defined and less bright compared to that of the iron oxide precursor (see panels b and h in Fig. 2), as likely due to a higher conductivity of the former as compared to the latter. A more defined view of the materials is given by the TEM, which suggests a pristine sample containing almost fully regular γ-$Fe_2O_3$ spherules with an approximate diameter from 50 to 90 nm (Fig. 2d). These particles are converted upon Ar-annealing into nano- (80 nm) and sub-micrometrical (500 nm) irregular α-Fe domains enclosed in a thin carbon shell (Fig. 2g). This morphology is retained after the final oxidation step to obtain α-$Fe_2O_3$@C (Fig. 2j). On the other hand, the EDS analyses actually reveal the complete reduction of the maghemite precursor (Fig. 2c) to metallic iron along with the formation of carbon with a weight ratio of 23% during pyrolysis (Fig. 2f), which is lowered to about 6% after the final oxidation of the intermediate to α-$Fe_2O_3$@C (Fig. 2i), in which all elements are homogeneously distributed (see corresponding map in inset of Fig. 2i).

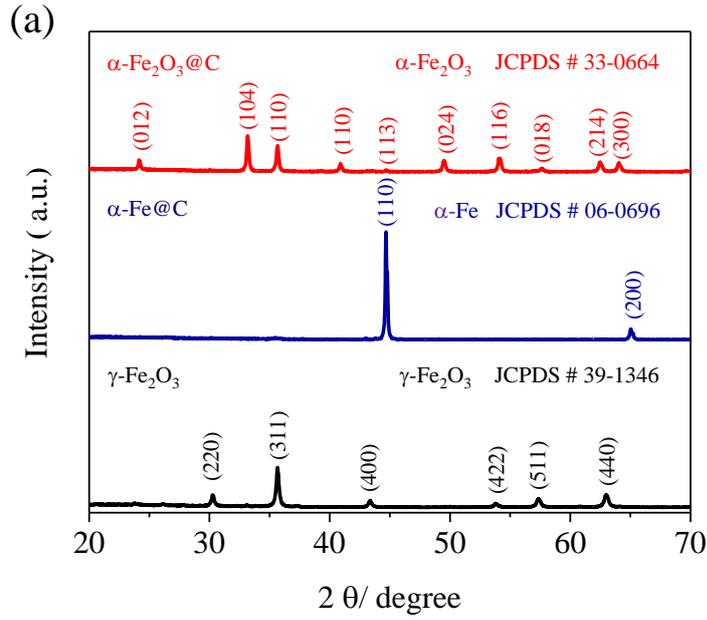

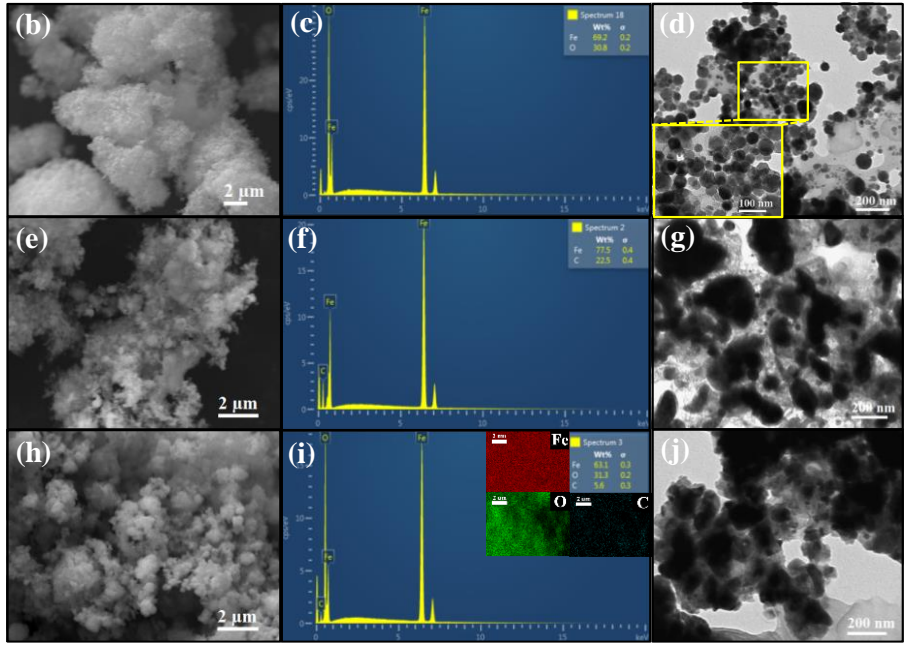

**Figure 2.** (**a**) XRD patterns of γ-Fe$_2$O$_3$, α-Fe@C and α-Fe$_2$O$_3$@C. (**b,e,h**) SEM, (**c,f,i**) EDS and (**d,g,j**) TEM of γ-Fe$_2$O$_3$ (**b-d**), α-Fe@C (**e-g**) and α-Fe$_2$O$_3$@C (**h-j**); Inset in (**i**) represents the elemental mapping of Fe, O and C of the α-Fe$_2$O$_3$@C.

According to our previous report on a conversion electrode based on nickel oxide,[35] we may expect that the morphology observed for α-Fe$_2$O$_3$@C in Fig. 2 would ensure optimal electric contact between the α-Fe$_2$O$_3$ grains and buffer volume variation, thus allowing an improved electrochemical process in Li-cell.[46] In addition, the encapsulation of nanoparticles into aggregates with a micro-size can actually ensure suitable electrode tap density and, at the same time, mitigate the electrolyte decomposition and increase the cell efficiency.[11,47]

The conversion process of α-Fe$_2$O$_3$@C is characterized by collecting CV and EIS data (see Figure 3). The CV of the γ-Fe$_2$O$_3$ precursor, which is reported for comparison (Fig. 3a), shows during the 1$^{st}$ discharge a small reduction peak occurring at about 0.9 V *vs* Li$^+$/Li, followed by an intense peak at about 0.7 V *vs* Li$^+$/Li and a shoulder at low voltage values. Instead, α-Fe$_2$O$_3$@C (Fig. 3c) shows one discharge process at ca. 0.7 V *vs* Li$^+$/Li and the low voltage slope. Hence, the cathodic response of the precursor reflects an initial Li$^+$ insertion into the γ-Fe$_2$O$_3$ structure at low lithiation degrees (peak at 0.94 V *vs* Li$^+$/Li in Fig. 2a),[48] the subsequent displacement process at 0.68 *vs* Li$^+$/Li of the oxide to metallic iron enclosed in a Li$_2$O matrix,[49] and the electrolyte decomposition leading to precipitation of the solid electrolyte interphase (SEI) at low potential values.[11] Interestingly, α-Fe$_2$O$_3$@C shows in the first cathodic scan only the displacement process and concomitant SEI formation at 0.69 V *vs* Li$^+$/Li (Fig. 3c).

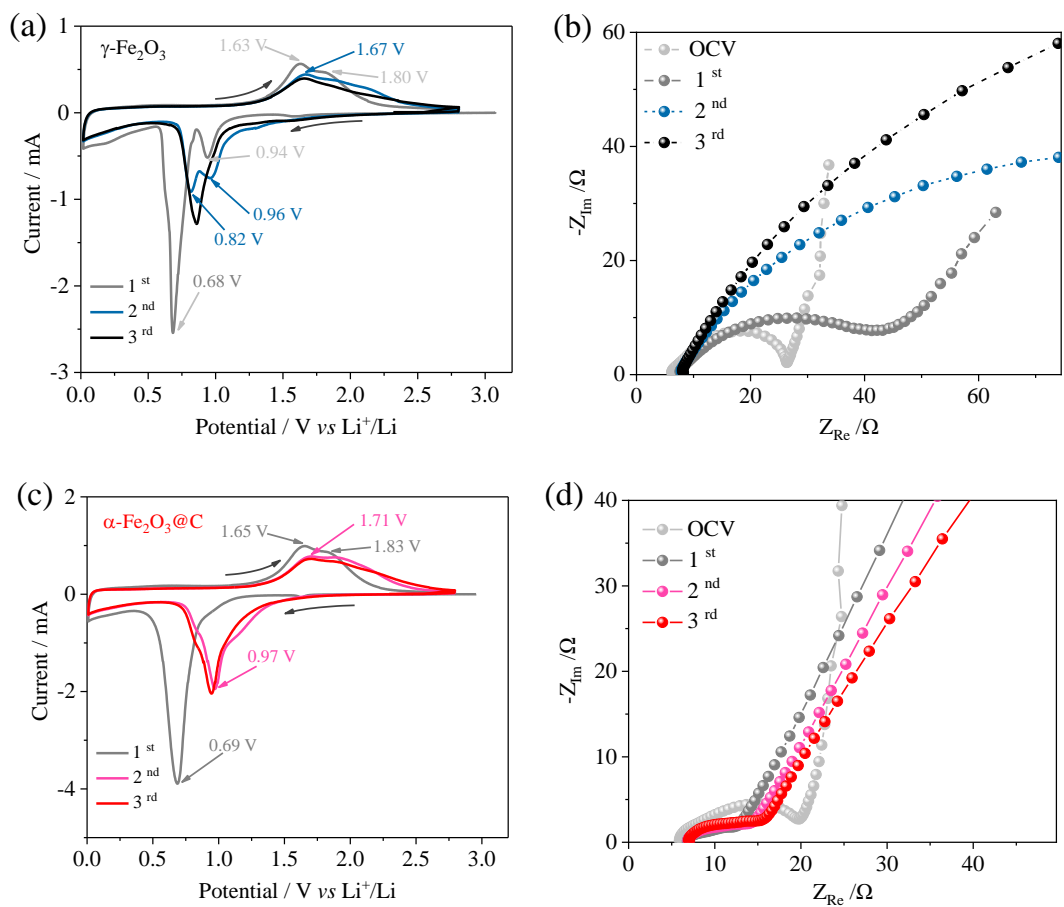

**Figure 3**. (**a,c**) CV profiles and (**b,d**) EIS Nyquist plots in the OCV condition and after 1, 2 and 3 cycles of γ-Fe$_2$O$_3$ precursor (**a,b**) and (**c,d**) α-Fe$_2$O$_3$@C composite in T-type cells with lithium disks as counter/reference electrodes and 1M LiPF$_6$ in EC:DMC (1:1, v/v) electrolyte. CV data collected from 2.8 to 0.01 V vs. Li$^+$/Li using a scan rate of 0.1 mV s$^{-1}$. Room temperature (25°C).

During the first anodic scan, both γ-Fe$_2$O$_3$ precursor and α-Fe$_2$O$_3$@C electrode have a similar behavior, as they show a broad wave at ca. 1.6 V *vs* Li$^+$/Li and a shoulder at ca. 1.8 V *vs* Li$^+$/Li, corresponding to the partially reversible conversion process [49]. The second voltammetry cycle reveals a gradual shift in potential for the reduction processes, namely to 0.96 V and 0.82 V *vs* Li$^+$/Li for γ-Fe$_2$O$_3$ (Fig. 3a) and to 0.97 V *vs* Li$^+$/Li for α-Fe$_2$O$_3$@C (Fig. 3c). On the other hand, during the subsequent cycles γ-Fe$_2$O$_3$ and α-Fe$_2$O$_3$@C undergo different trends, as also suggested by EIS (Fig. 3b and 3d). Accordingly, γ-Fe$_2$O$_3$ shows a gradual electrochemical deactivation (Fig. 3a) as well as a notable increase in electrode/electrolyte interface resistance, whose value may be extracted from the impedance data at high-to-middle frequency (Fig. 3b). Instead, α-Fe$_2$O$_3$@C shows a steady-state behavior (Fig. 3c), characterized by conversion processes stably leading to intense peaks at about 1.0 V *vs* Li$^+$/Li upon discharging and at about 1.7 V *vs* Li$^+$/Li upon charging, along with an EIS response without any sign of resistance increase (Fig. 3d). These observations suggest that the change in structure and morphology of the iron oxide powder, i.e., from γ-Fe$_2$O$_3$ to α-Fe$_2$O$_3$@C, greatly benefits the electrode charge transfer.

The α-Fe$_2$O$_3$@C is subsequently cycled in a lithium cell at a C-rate of C/5 (1C = 1007 mA g$^{-1}$) in comparison with the pristine oxide and the synthesis intermediate as reported in Figure 4 (panels a, b). The first discharge evolves mainly around 0.9 V with capacity exceeding the theoretical value for both γ-Fe$_2$O$_3$ and α-Fe$_2$O$_3$@C, that is, 1130 and 1370 mAh g$^{-1}$, respectively. The synthesis intermediate, which is formed by inactive iron and disordered carbon, delivers a capacity as low as 190 mAh g$^{-1}$ during the same reduction step, thereby suggesting that the excess capacity observed for the iron oxides could indeed be associated with the SEI formation,[11] as well as with possible contribution of the carbon matrix to the lithium exchange in the α-Fe$_2$O$_3$@C.[46] During charge process, the two iron oxides exhibit a sloped voltage signature over 1.7 V reflecting the typical charge/discharge hysteresis of the Li-conversion electrodes,[15,47] with first-cycle capacities of 840 mAh g$^{-1}$ for γ-Fe$_2$O$_3$ and 970 mAh g$^{-1}$ for α-Fe$_2$O$_3$@C. Notably, α-Fe$_2$O$_3$@C delivers a stable capacity upon the 80 cycles herein considered, with an average steady-state value of 860 mAh g$^{-1}$ and almost 100% Coulombic efficiency, while γ-Fe$_2$O$_3$ shows rapid capacity decay to values as low as 100 mAh g$^{-1}$ after only 20 cycles (see Fig. 4b).

The enhanced stability of the α-$Fe_2O_3$@C electrode compared to the pristine γ-$Fe_2O_3$ precursor may be reasonably attributed to an improved electric contact and structural retention upon the discharge/charge cycles provided by the carbon shell already observed in Fig. 2, as also suggested by our previous reports.[50] However, a further beneficial effect of the phase change from γ-$Fe_2O_3$ to α-$Fe_2O_3$ upon the thermal treatments cannot be excluded. In this regard, earlier works have shown that phase composition may have a significant effect on the behavior of $Fe_2O_3$ in the cell upon long term cycling.[51] Moreover, we remark that morphology certainly affects the electrochemical activity of conversion iron oxides.[52] The rate capability of the α-$Fe_2O_3$@C electrode is then investigated by performing a cycling test at currents ranging from C/10 to 2C (Fig. 4c, d). The panel d of Fig. 4 indicates a delivered capacity of about 900 mAh $g^{-1}$ at C/10, which decreases to about 800 mAh $g^{-1}$ when the current rate is increased to C/3 due to expected slight raise in charge/discharge polarization (Fig. 4c). A further increase in C rate lowers the reversible capacity to 430 mAh $g^{-1}$ at 2C, which is a remarkable value considering the high specific current (more than 2 A $g^{-1}$) as compared to the 2C rate of conventional graphite (0.744 A $g^{-1}$).[7] Furthermore, as the current rate is decreased back to the initial value of C/10 at the 45$^{th}$ cycle, the cell almost recovers the pristine capacity (i.e., about 900 mAh $g^{-1}$), thus accounting for a relevant stability of the α-$Fe_2O_3$@C material by changing the operating conditions. These results suggest that the synthesis pathway adopted herein allow the achievement of an improved α-$Fe_2O_3$@C conversion electrode that might be applied as the anode in a Li-ion battery.[22,35]

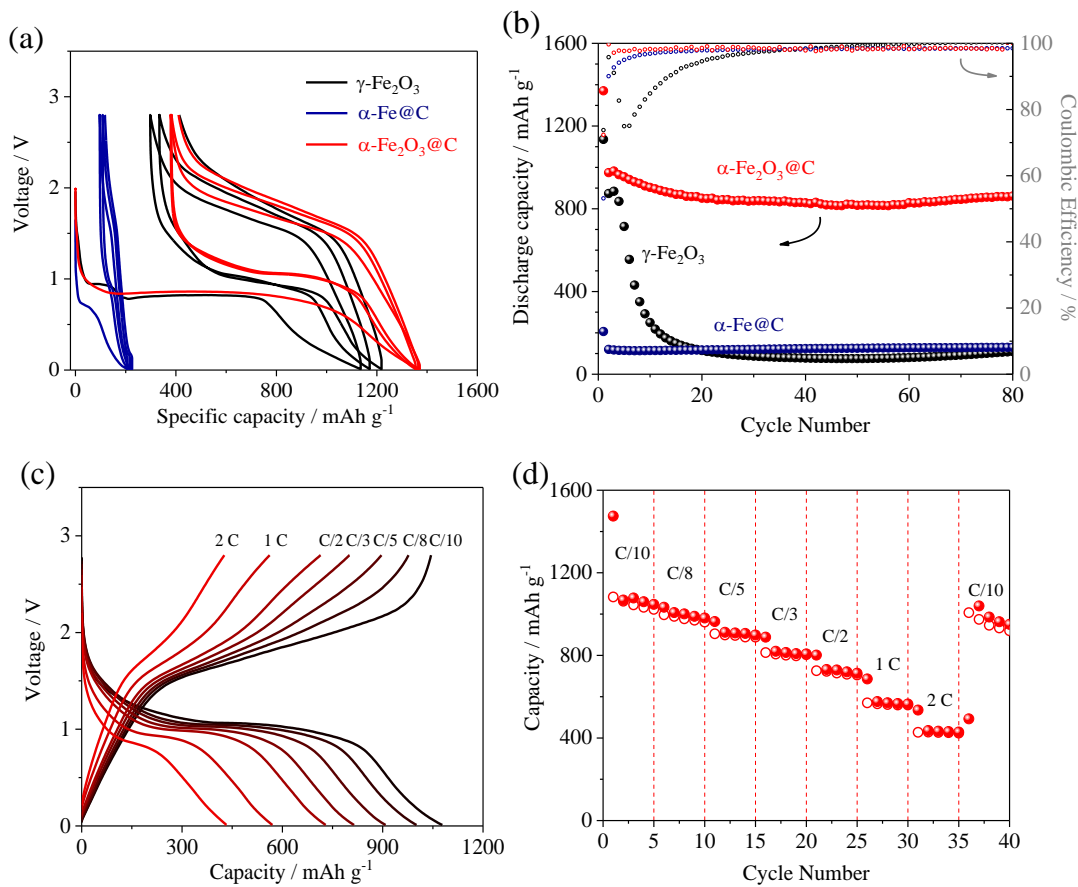

**Figure 4.** (**a**) Voltage curves and (**b**) trends of capacity and coulombic efficiency at a C/5 rate (1C = 1007 mA g$^{-1}$) of γ-Fe$_2$O$_3$ precursor, α-Fe@C intermediate, and α-Fe$_2$O$_3$@C composite in coin-type lithium half-cells. (**c**) Selected voltage curves and (**d**) trend of specific capacity in a rate capability test of the α-Fe$_2$O$_3$@C electrode in a coin-type lithium half-cell at C/10, C/8, C/5, C/3, C/2, 1C, and 2C rates (1C = 1007 mA g$^{-1}$). Electrolyte: 1M LiPF$_6$ in EC:DMC (1:1, v/v). Voltage range: 2.8 – 0.01 V. Room temperature (25°C).

A LiNi$_{0.5}$Mn$_{1.5}$O$_4$ cathode operating in lithium cell at a voltage as high as 4.8 V is synthesized (Figure 5) and considered the most adequate candidate for achieving a practical application of the Fe$_2$O$_3$@C conversion anode in a Li-ion battery.[36–38] The XRD pattern of the cathode powder reported in Fig. 5a reveals all the diffraction peaks characteristic of the $P4_332$ spinel phase (ICSD # 230819), without signs of impurities such as rock-salts (Li$_x$Ni$_{1-x}$O) causing deviation from the stoichiometric LiNi$_{0.5}$Mn$_{1.5}$O$_4$ composition in $Fd\bar{3}m$ phases.[53] Furthermore, Rietveld refinement of the XRD indicates a crystallite size of about 100 nm, which is confirmed by the morphological details detected by SEM (Fig. 5b). Indeed, the SEM shows nano-sized primary particles agglomerated into domains with a micro-size, which is considered a suitable configuration for ensuring, at the same time, improved electrochemical activity and limited parasitic electrolyte decomposition.[54] Nanoparticles can in fact improve the (de)insertion kinetics of the lithium into and from the spinel structure of the cathode, and strengthen concomitantly the electrolyte oxidation at high-voltage due to their relatively high specific surface.[55] On the other hand, the latter side process may be remarkably suppressed by agglomerating the nanoparticles into microcrystals suitable for achieving high efficiency in battery of the LiNi$_{0.5}$Mn$_{1.5}$O$_4$ cathode.[56]

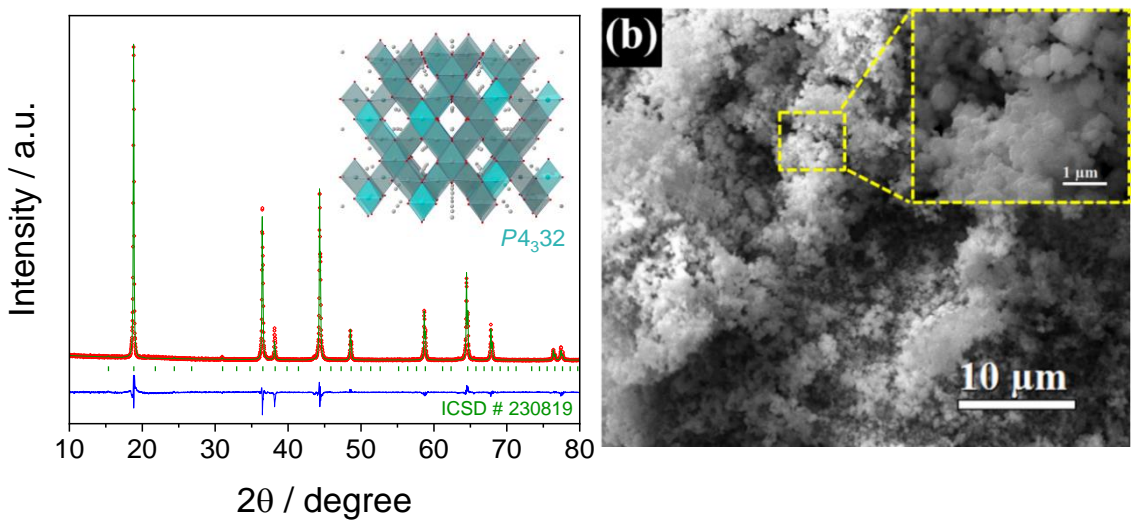

**Figure 5.** (**a**) Experimental and refined XRD pattern with reference (ICSD # 230819) and (**b**) SEM image of the LiNi$_{0.5}$Mn$_{1.5}$O$_4$ sample with magnification in inset.

The electrochemical response of LiNi$_{0.5}$Mn$_{1.5}$O$_4$ is preliminary investigated in a lithium half-cell by performing a cycling test at C/2 rate (1 C = 147 mA g$^{-1}$), as reported in Figure 6a, b. The cathode shows a stable specific capacity of 120 mAh g$^{-1}$ as well as a coulombic efficiency approaching 100% after few charge-discharge cycles (Fig. 6a). We remark herein that LiNi$_{0.5}$Mn$_{1.5}$O$_4$ may have either a disordered (space group $Fd\bar{3}m$) or an ordered structure (space group $P4_332$) depending on the presence of oxygen deficiency caused by the above-mentioned rock-salt impurities. In the first arrangement Ni$^{2+}$ and Mn$^{4+}$ ions are evenly positioned in the 16d and the 8a sites, whilst in the latter one Ni$^{2+}$, Mn$^{4+}$, and Li$^+$ ions are respectively distributed in the 4b, 12d, and 8c crystal positions. Notably, Mn$^{3+}$ occurs in the $Fd\bar{3}m$ framework to balance the oxygen deficiency, thereby leading to a characteristic voltage profile showing a further process at ca. 4.0 V *vs* Li$^+$/Li.[53] In this regard, the half-cell voltage profile (Fig. 6b) reveals during the first charge an irreversible plateau at about 4.2 V, most likely attributed to the oxidation of water traces into the electrolyte or undetected Mn$^{3+}$ in the LiNi$_{0.5}$Mn$_{1.5}$O$_4$ electrode[53,57], as well as possible Al corrosion.[58] Furthermore, a capacity excess at high voltages during the first cycle suggests the occurrence of electrolyte decomposition with deposition of a suitable SEI film which protects the electrode surface.[59] Accordingly, the subsequent cycles reveal only the typical high voltage signature to the Ni$^{4+}$/Ni$^{2+}$ redox pair, that is, around 4.7 − 4.8 V.[60]

It is worth mentioning that LiNi$_{0.5}$Mn$_{1.5}$O$_4$-based batteries typically suffer from poor stability at moderately elevated temperature, showing degradation of the electrode surface and manganese dissolution in the electrolyte solution. Coating LiNi$_{0.5}$Mn$_{1.5}$O$_4$ with thin oxide layers of various nature, such as ZnO, TiO$_2$, and Al$_2$O$_3$, has proven to effectually stabilize the electrode/electrolyte interphase and improve the high-temperature response of the cell.[61–63] Furthermore, we have recently demonstrated that slight changes in formulation of the oxalate precursors during the modified co-precipitation synthesis of LiNi$_{0.5}$Mn$_{1.5}$O$_4$ may produce a multi-metal spinel framework with enhanced behavior at 55 °C.[53] These strategies might be suitable for enabling application of the $Fe_2O_3$@C/$LiNi_{0.5}Mn_{1.5}O_4$ battery above room temperature. The characteristics of this full cell are displayed in Figure 6, showing the voltage curves (Fig. 6c) and the trends of capacity and Coulombic efficiency (Fig. 6d) by referring the specific values to the cathode mass. The α-$Fe_2O_3$@C/$LiNi_{0.5}Mn_{1.5}O_4$ full cell cycled at the constant current rate of C/2 (1C = 147 mA g$^{-1}$) exhibits a stable response, delivering ca. 115 mAh g$_{cathode}^{-1}$ during the initial stages of the test and ca. 110 mAh g$_{cathode}^{-1}$ upon the subsequent cycles, with a columbic efficiency approaching 97% (Fig. 6c). The related voltage profile during the first cycle (inset of Fig. 6d) show an irreversible profile, possibly ascribed to the above discussed side processes, that is, structural reorganizations, electrolyte decomposition and side reactions with SEI film formation both on the anode and on the cathode.[53,57–59] Besides, the voltage profiles at the 2$^{nd}$, 5$^{th}$, 10$^{th}$, 20$^{th}$, 30$^{th}$ and 60$^{th}$ cycle (Fig. 6d) suggest a steady-state sloped signal centered around 3.2 V, reflecting the conversion process of the α-$Fe_2O_3$@C anode[47] and the simultaneous (de)insertion of the $LiNi_{0.5}Mn_{1.5}O_4$ cathode,[44] that reversibly occur during full Li-ion cell operation.[60] Taking into account the capacity and average voltage values shown in Figure 6 (c and d), i.e., 110 mAh g$^{-1}$ and 3.2 V, respectively, we can estimate for the $Fe_2O_3$@C/$LiNi_{0.5}Mn_{1.5}O_4$ full cell a theoretical gravimetric energy density of about 350 Wh kg$^{-1}$, and a practical value approaching 150 Wh kg$^{-1}$. Therefore, this battery configuration might outperform similar systems employing a cheap iron oxide conversion anode and cobalt-free spinel cathode.[33] On the other hand, the conventional graphite/$LiFePO_4$ battery may ensure a capacity as high as 150 mAh g$^{-1}$ (when referred to the cathode mass) and operates at ca. 3.2 V, thus having a theoretical energy of 480 Wh kg$^{-1}$, which might be reflected as a practical value of more than 160 Wh kg$^{-1}$.[64] Notably, this latter system possesses even more interesting features in term of environmental friendliness and cost due to the absence of cobalt and nickel in the electrode formulation, although the low density of olivine materials adversely affects the volumetric energy of the whole battery.[65]

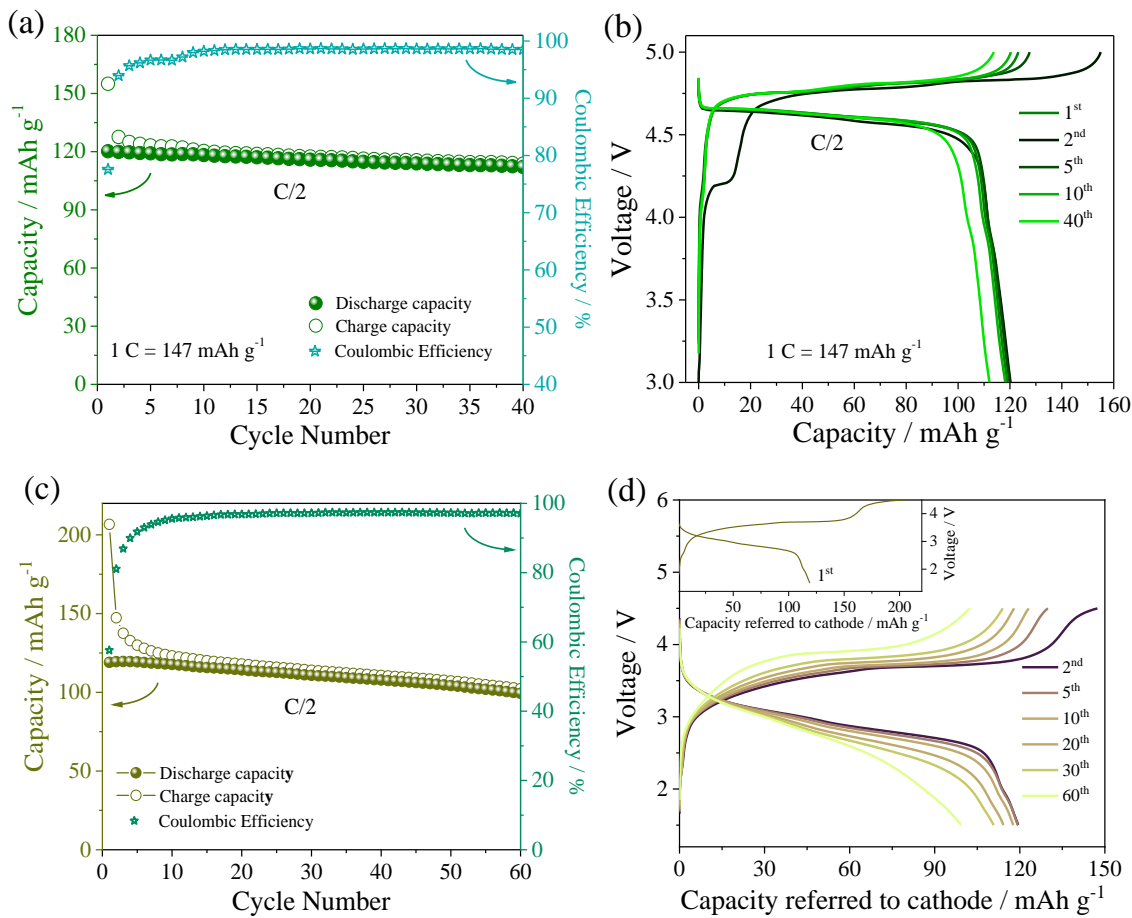

**Figure 6**. (**a**) Trends of specific capacity and coulombic efficiency (**b**) and selected voltage curves of the LiNi$_{0.5}$Mn$_{1.5}$O$_4$ electrode in coin-type lithium half-cell at a C/2 rate (1C =147 mA g$^{-1}$) within 3.0 V and 5.0 V. (**c** and **d**) Electrochemical response within 1.5 V – 4.5 V range of the α-Fe$_2$O$_3$@C/LiNi$_{0.5}$Mn$_{1.5}$O$_4$ full-cell at a C/2 rate referred to the LiNi$_{0.5}$Mn$_{1.5}$O$_4$ cathode mass (1C =147 mA g$^{-1}$) in terms of (**c**) trends of specific capacity and coulombic efficiency and (**d**) selected voltage profiles (1$^{st}$ cycle in inset). Electrolyte: LiPF$_6$ 1M in EC:DMC (1:1, v/v). Voltage range: 2.8 – 0.01 V. Room temperature (25°C).

**Conclusions**

An alternative α-Fe$_2$O$_3$@C conversion anode and a high-voltage LiNi$_{0.5}$Mn$_{1.5}$O$_4$ spinel cathode have been synthesized, investigated, and combined in a new Li-ion cell. The anode has been synthesized by a two-step pathway leading to nanoparticles with a uniform and thin carbon shell, whilst the cathode has been prepared by an alternative approach as nanoparticles agglomerated into micro-domains. The α-Fe$_2$O$_3$@C nanocomposite has exhibited an excellent electrochemical performance, delivering a capacity that approaches 900 mAh g$^{-1}$, with a stable cycling trend and a suitable rate capability. This simple, fast, and cheap synthesis has been therefore suggested as an advantageous and scalable pathway as for obtaining high-capacity conversion metal oxides for battery application. Moreover, the LiNi$_{0.5}$Mn$_{1.5}$O$_4$ cathode has revealed well-suited structure and morphology as well as an adequate cycling response in the lithium half-cell, that is, a specific capacity exceeding 120 mAh g$^{-1}$, a high efficiency upon the first cycle, and an operating voltage of 4.8 V due to the redox activity of Ni$^{4+}$/Ni$^{2+}$. Hence, the combination of the α-Fe$_2$O$_3$@C and LiNi$_{0.5}$Mn$_{1.5}$O$_4$ electrodes has led to a full cell which may work at a C/2 rate (referred to mass of cathode). This new Li-ion cell has delivered a stable capacity of 110 mAh g$_{cathode}^{-1}$ at ca. 3.2 V with a satisfactory coulombic efficiency (higher than 97%). Therefore, we have estimated a practical energy density approaching 150 Wh kg$^{-1}$ and proposed the α-Fe$_2$O$_3$@C/LiNi$_{0.5}$Mn$_{1.5}$O$_4$ battery as low-cost energy storage system with modest environmental impact.

**Acknowledgements**

This work has received funding from the European Union's Horizon 2020 research and innovation programme Graphene Flagship under grant agreement No 881603 and the grant "Fondo di Ateneo per la Ricerca Locale (FAR) 2020", University of Ferrara, and has been performed within the collaboration project "Accordo di Collaborazione Quadro 2015" between University of Ferrara (Department of Chemical and Pharmaceutical Sciences) and Sapienza University of Rome (Department of Chemistry).

**Table of content**

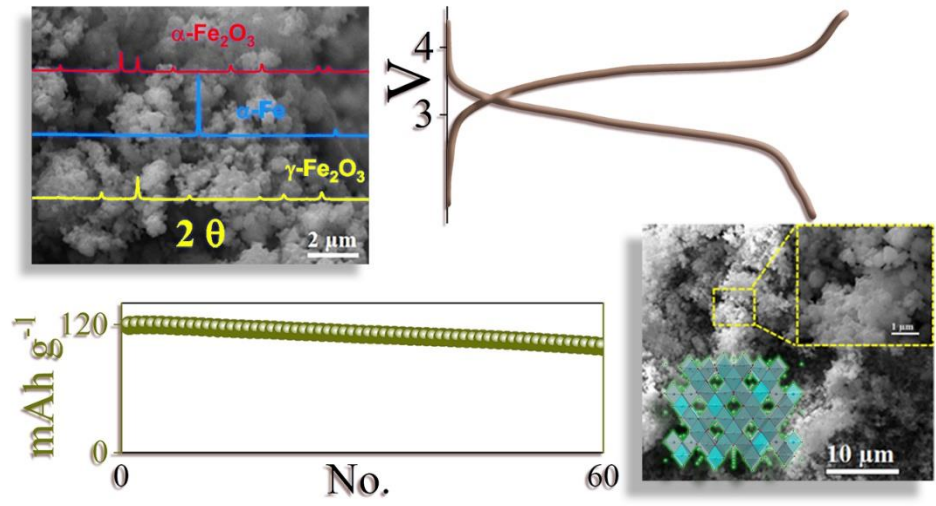